\documentclass[10pt]{article}
\usepackage{epsfig,graphicx,amssymb,amsfonts}
\usepackage{latexsym}
\newcommand{\be}{\begin{equation}}
\newcommand{\ee}{\end{equation}}
\newcommand{\bea}{\begin{eqnarray}}
\newcommand{\eea}{\end{eqnarray}}

\usepackage{cite}
\usepackage{epsfig}
\usepackage{amsfonts}

\def \beq {\begin{equation}}
\def \eeq {\end{equation}}
\def \bea {\begin{eqnarray}}
\def \eea {\end{eqnarray}}

 \def\e{{\rm e}}

\def\Z#1{_{\lower2pt\hbox{$\scriptstyle#1$}}}
\def\X#1{_{\lower2pt\hbox{$\scriptscriptstyle#1$}}}

\def\ApJ#1{Astrophys.\ J.\ {\bf#1}}

\begin{document}
\baselineskip=20pt

\vspace*{0cm}

\begin{flushright}
0709.3096 [hep-th]
\end{flushright}

\begin{center}
{\Large{\bf Remarks on Dynamical Dark Energy Measured by \\the
Conformal Age of the Universe }}

\vspace*{0.3in} Ishwaree P.\ Neupane \vspace*{0.3in}

\it Department of Physics and Astronomy, University of Canterbury\\
Private Bag 4800, Christchurch 8020, New Zealand\\
{\tt ishwaree.neupane@canterbury.ac.nz} \\
\vspace*{0.2in}
\end{center}

\begin{abstract}

We elaborate on a model of conformal dark energy (dynamical dark
energy measured by the conformal age of the universe) recently
proposed in [H. Wei and R.G. Cai, arXiv:0708.0884] where the
present-day dark energy density was taken to be $\rho_q \equiv 3
\alpha^2 m_P^2/\eta^2$, where $\eta$ is the conformal time and
$\alpha$ is a numerical constant. In the absence of an interaction
between the ordinary matter and dark energy field $q$, the model
may be adjusted to the present values of the dark energy density
fraction $\Omega\Z{q} \simeq 0.73$ and the equation of state
parameter $w\Z{q} < -0.78$, if the numerical constant $\alpha$
takes a reasonably large value, $\alpha\gtrsim 2.6$. However, in
the presence of a nontrivial gravitational coupling of $q$-field
to matter, say $\widetilde{Q}$, the model may be adjusted to the
values $\Omega\Z{q}\simeq 0.73$ and $w\Z{q}\simeq -1$, even if
$\alpha\sim {\cal O}(1)$, given that the present value of
$\widetilde{Q}$ is large. Unlike for the model in [R.G. Cai,
arXiv:0707.4049], the bound $\Omega\Z{q} <0.1$ during big bang
nucleosynthesis (BBN) may be satisfied for almost any value of
$\alpha$. Here we discuss some other limitations of this proposal
as a viable dark energy model. The model draws some parallels with
the holographic dark energy; we also briefly comment on the latter
model.

\bigskip \noindent {PACS numbers: 95.36.+x, 98.80.Es}

\end{abstract}

\vfill
\newpage
\baselineskip=20pt

\section{Introduction}

Inflation is an attractive paradigm for explaining small
temperature fluctuations in the cosmic microwave background, the
distribution of galaxies, the homogeneity and isotropy of the
universe on scales of more than $100$~Mpc and its spatial
flatness, as inferred by recent WMAP data~\cite{WMAP}. The current
standard model of cosmology somehow combines the original hot big
bang model and the primordial inflation~\cite{Linde:87}, by virtue
of the existence of a fundamental scalar field, called {\it
inflaton}. However, the standard model of cosmology has some gaps
and cracks; for instance, the recently observed accelerated
expansion of the universe~\cite{supernovae} appears to suggest in
the fabric of the cosmos a self-repulsive dark energy component of
magnitude about $73\%$ of the total energy budget of the entire
universe. Evidence in favour of this accelerated expansion has
strengthened significantly as the result of further SNe Ia
observations~\cite{Astier:05}, surveys of large scale
structure~\cite{Tegmark:SDSS} and improved measurements of the
cosmic microwave background~\cite{Spergel:06}. The precise cause
of this late-time acceleration and the nature of dark energy
attributed to this effect, however, remain illusive.

The phenomenal role of a cosmological vacuum energy (or dark
energy) has changed our vocabulary for describing the cosmological
possibilities and the fate of our universe
(see~\cite{Weinberg:1988,Padmanabhan} for reviews). We do not
understand whether the highly accelerated expansion shortly after
the big-bang - called inflation and the current accelerated
expansion of the universe (caused by dark energy) are related.
Understanding of dark energy's origin may be expected to provide
some useful insights to many other puzzles in physics, including:
What caused the early universe inflation? Why does dark
energy/dark matter make up most of the universe?

In a fundamental theory of gravity plus elementary particles and
fields, it is quite plausible that the primordial inflation
naturally led to have a dark energy effect in the conditions of
concurrent universe, i.e. when the universe became much larger
than its size at the beginning. Such an effect can be explained
through two somehow different mechanisms. In the first, and
perhaps the most viable
approach~\cite{Ratra:88,Peebles:99a,Ish:07b}, the present-day dark
energy effect could be realized as a remnant of the original
inflaton field that went into a hide shortly after reheating (or
even after inflation), but which started to play a new role during
the matter dominated epoch, especially, on large cosmological
scales ($> 100$ Mpc), where gravity would almost fail to curve the
spacetime, thereby leading to a spatially flat
Friedmann-Lama\^itre-Robertson-Walker (FLRW) universe. In the
second approach, the quantum fluctuations associated with an
accelerating slice of a FRW metric (during the primordial
inflation) could gradually overtake at late times the ambient
matter distributions, tending to increase the rate of expansion of
the universe on large cosmological scales. In this paper we
discuss about the latter possibility, in the framework of a model
of ``conformal" dark energy (dynamical dark energy measured by the
conformal age of the universe) recently proposed by Cai and
Wei~\cite{Wei:07a}.

There has also been a fair amount of interest in the possibility
that the dark energy is holographic~\cite{Cohen:1998,Hsu:2004}.
The model of dynamical dark energy discussed in~\cite{Wei:07a} has
some similarities with the so-called holographic dark energy
proposed earlier by Li~\cite{Li:2004}. We will briefly comment at
the end on holographic dark energy models.

\section{Why scalar gravity after all?}

The possibility remains that the cosmological constant (or the
vacuum energy) is fundamentally variable. In order to give the
idea a fair hearing, one should conceivably take some sort of dark
energy potential. An appropriate Lagrangian might be
\begin{equation}\label{action}
{\cal L} = \sqrt{-g} \left( \frac{R}{2\kappa^2} -\frac{1}{2}
(\partial q)^2- V(q) \right)+ {\cal L}_{m},
\end{equation}
where $\kappa$ is the inverse Planck mass $m\Z{P}^{-1}=(8\pi
G\Z{N})^{1/2}$, $G\Z{N}$ is Newton's constant, $q$ is a
fundamental scalar (or dark energy) field and $V(q)$ is its
potential. Indeed, in the simplest dynamical dark energy
models~\cite{Ratra:88}, dark energy is associated with the energy
density of a scalar field with a canonical kinetic structure, as
above. Most dynamical dark energy models, including the
``agegraphic" (actually, inverse age-mapping) and holographic dark
energy, may be analyzed by maintaining the above structure of the
theory.

For an analytic treatment it is necessary to evaluate the
equations generated by variation of the action~(\ref{action});
thus a particular choice of a metric has to be made. In line with
current observations, and because it greatly simplifies the
calculations, we make the rather standard choice of a spatially
flat, homogeneous metric: $ds^2 = -dt^2 + a^2(t)\, d {\bf x}^2$,
where $a(t)$ is the scale factor of a spatially flat FRW universe.
This is consistent with the measurements of the cosmic microwave
background (CMB) anisotropies and large-scale structures of the
universe, which indicate that the present universe is spatially
flat and homogeneous on large scales.

An important ingredient of a cosmological model is matter
Lagrangian, which may be given by~\cite{Damour:1993id}
\begin{equation}\label{matter-scalar}
{\cal L}_m\equiv {\cal L} (\beta^2(q) g_{\mu\nu},\psi_m)=
\sqrt{-g}\, \beta^4(q) \,\widetilde{\rho}\Z{i},
\end{equation}
where $\widetilde{\rho}\Z{(i)} \propto
{\hat{a}}^{-3\left(1+w_i\right)}$ ($i=$m, r), $\hat{a}\equiv a
\beta(q)$. Introduction of a fundamental scalar field $q$, its
potential $V(q)$ and the coupling $\beta(q)$ between $q$ and the
ordinary matter ($\rho\Z{m}$) and radiation ($\rho\Z{r}$) may not
be arbitrary rather a requirement for the present-day concordance
model cosmology. These ingredients are strongly motivated by
supergravity and superstring theories.

Einstein's equations following from Eqs. (\ref{action}) and
(\ref{matter-scalar}) are~\cite{Ish:07e}
\begin{eqnarray}
3 H^2 &=& \kappa^2 \left( \frac{1}{2}\, \dot{q}\,^2 + V(q)
+\beta^4 \left(\rho_{m}+\rho_{r}\right) \right),\label{non-minimal1}\\
- 2 \dot{H} &=& \kappa^2 \left( \dot{q}\,^2  + \beta^4 \left(1+
w_m\right) \rho_{m}  +\frac{4}{3} \beta^4 \rho_r\right),
\label{non-minimal2}
\end{eqnarray}
where $w_i\equiv p_{i}/\rho_{i}$ and $\rho_{i} \propto
(a\beta)^{-\,3(1+w_{i})}$. The scalar field $q$ couples to the
trace of the matter stress tensor, $g_{(i)}^{\mu\mu}
T_{\mu\nu}^{(i)}$, namely
\begin{equation}
-\nabla^2 q= \ddot{q}+ 3H \dot{q} = -V_{ ,\, q} + \alpha\Z{q}
T\Z{\mu (i)}^\mu,
\end{equation}
where $\alpha\Z{q} \equiv \frac{d\ln \beta(q)}{dq} $ and $H \equiv
\dot{a}/a$ is the Hubble parameter (the dot denotes a derivative
with respect to cosmic time $t$). Since $T\Z{\mu (m)}^\mu= -
\rho\Z{m}+ 3 p\Z{m} \equiv -\rho\Z{m}\left(1-3 w\Z{m}\right)$ and
$T\Z{\mu (r)}^\mu= - \rho\Z{r}+ 3 p\Z{r} =0$, the above equation
of motion for $q$ can be expressed in the following
form~\footnote{The parameter $\alpha\Z{q}$ defined here
corresponds to $-\,Q$ in refs.~\cite{Ish:07a,Ish:07e}, where it
was assumed that $Q<0$.}:
\begin{equation}
\dot{\rho}\Z{q} + 3 H \rho\Z{q} \left(1 + w\Z{q}\right) = -
\dot{q}\gamma \alpha\Z{q} \beta(q) \rho\Z{m},\label{nonminimal1}
\end{equation}
where $\gamma \equiv (1-3 w\Z{m})$, $w\Z{m}\equiv
p\Z{m}/\rho\Z{m}$, $\rho\Z{q} \equiv \frac{1}{2} \dot{q}^2 +
V(q)$, $w\Z{q}\equiv p\Z{q}/\rho\Z{q}$. This equation, along with
the equations of motion for ordinary fluids (matter and
radiation):
\begin{equation}
\dot{\rho}\Z{m} +3H \rho\Z{m} (1+ w\Z{m}) =  + \dot{q} \gamma
\alpha\Z{q} \beta(q) \rho\Z{m}, \quad \dot{\rho}\Z{r} +4H
\rho\Z{r} =0, \label{nonminimal2}
\end{equation}
guarantees the conservation of total energy, namely $\dot{\rho}+
3H (\rho + p)=0$, where $\rho\equiv \rho\Z{m}+\rho\Z{r}+
\rho\Z{q}$.

The set of autonomous equations of motion may be given by (see,
e.g.~\cite{Ish:07a,Ish:07e})
\begin{eqnarray} \Omega\Z{r}+ 3w\Z{q}
\Omega\Z{q}+ 3w\Z{m} \Omega\Z{m}
+2\varepsilon+3 &=&0,\label{EoM1}\\
\Omega\Z{q}^\prime+ 2\varepsilon \Omega\Z{q} +
3(1+w\Z{q})\Omega\Z{q} &=& - \widetilde{Q},\label{EoM2} \\
\Omega\Z{m}^\prime + 2\varepsilon \Omega\Z{m} +
3(1+w\Z{m})\Omega\Z{m} &=& + \widetilde{Q},\label{EoM3}
\end{eqnarray}
subject to the Friedmann constraint $\Omega_{m}+\Omega_{r}+
\Omega_{q}=1$, where the prime denotes the derivative with respect
to ${N}\equiv \ln [a(t)] +{\rm const}$, $\varepsilon=\dot{H}/H^2$,
$\widetilde{Q}\equiv \gamma q^{\,\prime} \alpha\Z{q} \Omega_{m}$,
$q^{\,\prime} ={\dot{q}}/{H}$, $\Omega_{i} \equiv \kappa^2
{\beta^4 \rho_{i}}/{(3H^2)}$ and $\Omega_{q}=\kappa^2
{\rho_{q}}/{(3H^2)}$. The fact that the radiation term $\rho\Z{r}$
does not contribute to the scalar potential or the Klein-Gordon
equation has an interesting implication: in the early universe,
e.g. during or shortly after inflation, one can ignore the
coupling $\beta(q)$, since $\rho\Z{m} \ll \rho\Z{r}$. During the
matter-dominated universe, given that $\rho\Z{m}\propto 1/a^3$,
$w\Z{m}\simeq 0$ and $a\propto t^{2/3}$ ($\varepsilon=-3/2$), it
is plausible that $\widetilde{Q} \approx 0$. However, the coupling
$\widetilde{Q}$ may be relevant especially when $\rho\Z{q}\gtrsim
\rho\Z{m}$, i.e., in the dark energy-dominated universe.

From Eq.~(\ref{EoM1}), we find that the dark energy equation of
state is given by
\begin{equation}\label{de-EoS}
w\Z{\rm DE}\equiv w_{q} = - \frac{2\varepsilon+3+ 3\sum_i
w_{i}\Omega_{i}+\Omega_{r}}{3\Omega_{q}},
\end{equation}
where $i=m$ (matter) includes all forms of matter fields, such as,
pressureless dust ($w=0$), stiff fluid ($w=1$), cosmic strings
($w=-1/3$), domain walls ($w=-2/3$), etc. One might also note the
universe accelerates when the effective equation of state $w_{\rm
eff}$ is less than $-1/3$ (where $w\Z{\rm eff} \equiv
-1-2\varepsilon/3$), not when $w_{q}<-1/3$. In the particular case
that $w\Z{m}= 0$ and $\Omega_{r}\approx 0$, so that the matter is
approximated by a pressureless non-relativistic perfect fluid, the
universe accelerates for
\begin{equation}
 w_{q} \Omega_{q} < - \frac{1}{3}.
\end{equation}
With the input $\Omega\Z{q}=0.73$, we can see that the universe
accelerates for $w\Z{q}< - \,0.46$.

\section{What is dark energy?}

We do not yet have any clue as to what dark energy is, and how to
compute its present contribution from the first principles. A
common lore is that ``dark energy" is the Einstein's cosmological
constant until proven otherwise, for the reason that it is the
most economical interpretation of the data. The main observation
that has led to this viewpoint is the following: the combination
of WMAP3 and Supernova Legacy Survey data sets show a significant
constraint on the dark energy equation of state, $w\Z{\rm
DE}=-0.97^{+0.07}_{-0.09}$, on the $\Lambda$CDM model, i.e., in a
flat universe, with a prior $w\Z{m}=0$. Perhaps this observation
is not yet sufficiently convincing to abandon other possibilities,
at least, for two other reasons: firstly, no theoretical model,
not even the most sophisticated, such as supersymmetry or string
theory, is able to explain the presence of a small positive
cosmological constant, in the amount that our observations
require~\cite{Weinberg:1988}, $\rho_\Lambda\sim 5\times
10^{-27}~{\rm kg/m^3}$ or $\rho_\Lambda \sim 10^{-123}$ in Planck
units; secondly, there are widespread claims that the analysis of
the type Ia supernova data sets actually favour a time-varying
dark energy equation of state at higher redshifts (see,
e.g.~\cite{Sahni}, for a review).

Needless to emphasize, the possibility remains that dark energy is
fundamentally variable. It is thus a fair approach to envisage for
plausible phenomenological models and apply the observational
results either to rule them all or select one of them. In order to
give the idea a fair treatment, in this work we briefly review
some recent attempts in this direction, namely, the models of
conformal and ``holographic" dark energy.

\subsection{Dark energy measured by a cosmic time}

In a recent proposal~\cite{Cai:07a}, Cai argued that the
present-day dark energy density may be defined by the energy
density of metric fluctuations in a Minkowski spacetime, namely
\begin{equation}\label{quantum-de}
\rho\Z{\Lambda} \equiv \rho\Z{q} \propto \frac{1}{t^2\Z{P} t^2}
\equiv \frac{3 n^2\,m_{p}^2}{t^2},
\end{equation}
where the numerical coefficient $n \sim {\cal O}(1)$ and $t_{P}$
is Planck's time. The above relation is somehow based on quantum
kinematics or Heisenberg uncertainty type relations that put a
limit on the accuracy of quantum measurements; we refer to the
papers~\cite{Sasakura,Maziashvili} and references therein, for
further details. Without any reference to the field potential
$V(q)$, by Eq.~(\ref{quantum-de}), one can perhaps understand that
the quantum fluctuations in a Minkowski spacetime contribute to
the expectation value of the stress tensor in a way that mimics
the dark energy density at the present epoch. According
to~\cite{Cai:07a}, the cosmic time
\begin{equation}\label{age-universe} t=\int_{0}^a
\frac{da}{H\,a}=\int H^{-1} d\ln a
\end{equation}
may be considered as the age of our universe. Differentiating this
equation with respect to $\ln {a}$, we get
\begin{equation}
\frac{d t}{d\ln a}= \frac{1}{H}.
\end{equation}
Further, from the definition
\begin{equation}\label{def-Omega-q}
\Omega\Z{q}\equiv \frac{\rho\Z{q}}{3 m\Z{P}^2 H^2}=\frac{n^2}{t^2
H^2},
\end{equation}
we get
\begin{equation}\label{t-H-relation1}
t H =\pm \frac{n}{\sqrt{\Omega\Z{q}}}
\end{equation}
With $n>0$, because of the requirement that $t H>0$, we shall take
the positive sign in (\ref{t-H-relation1}). Then, differentiating
Eq.~(\ref{def-Omega-q}) with respect to $\ln a$, we get
\begin{equation}\label{constr-old}
\Omega\Z{q}^\prime +2 \varepsilon\Omega\Z{q} + \frac{2}{n}
\left(\Omega\Z{q}\right)^{3/2} =0.
\end{equation}
In the absence of interaction between the $q$-field and matter, so
that $\widetilde{Q}=0$, from Eq.~(\ref{EoM2}), we find
\begin{equation}\label{wq-EoS1-old}
 w\Z{q} =
-1-\frac{1}{3}\frac{\Omega\Z{q}^\prime}{\Omega\Z{q}}-\frac{2\varepsilon}{3}.
\end{equation}
Comparing Eqs.~(\ref{constr-old}) and (\ref{EoM2}) we get
\begin{equation}\label{wq-constr-old}
w_{q}= - 1 + \frac{2}{3\,n}\, \sqrt{\Omega\Z{q}}.
\end{equation}
Obviously, with $\sqrt{\Omega\Z{q}}/n>0$, or $t H>0$, we get
$w\Z{q}> -1$, in which case $q$ behaves as a canonical scalar
field or quintessence. From (\ref{wq-constr-old}) it is easy to
see that the $q$-field violates the strong energy condition,
$w\Z{q}\ge -1/3$, for $\sqrt{\Omega\Z{q}}< n$, which is the
minimal condition for a cosmic acceleration to occur in the
absence of ordinary fluids (matter and radiation). With the input
$\Omega\Z{q}= 0.73$, $w\Z{q}<-1/3$ for $n> 0.85$. The WMAP
observations, which are sensitive to $w\Z{q}$ over a redshift
range of roughly $1100$ (since decoupling), imply $w\Z{q}< -0.78$
($95\%$ confidence level), which translates to the condition $n>
3\, \sqrt{\Omega\Z{q}}$. This last condition obviously leads to a
result consistent with the discussion in~\cite{Wu:07a}, where the
best fit values were found to be $n=3.4$ and $\Omega\Z{q}=0.72$ in
using the constraints from CMB and LSS observations.

Equating Eqs.~(\ref{wq-EoS1-old}) and (\ref{wq-constr-old}) and
then solving for $\Omega\Z{q}$, we obtain
\begin{equation}\label{main-soln}
\frac{n}{\sqrt{\Omega\Z{q}}}=  \left\{\begin{array}{l} \frac{1}{2}
\left( 1+
b\Z{1}\,a^{-2} \right) \qquad ({\rm RD}, \quad a(t)= a\Z{{\rm r}, {\rm ini}}\,
t^{1/2}), \\ \\
 {2\over 3} \left( 1 + b\Z{2}\,a^{-3/2} \right)
\qquad ({\rm MD}, \quad a(t)= a\Z{{\rm m}, {\rm ini}}\, t^{2/3}),
\end{array} \right.
\end{equation}
where $b\Z{1} $ and $b\Z{2}$ are integration constants, and
$a\Z{{\rm r,~ ini}}$ and $a\Z{{\rm m,~ ini}}$ are scale factors at
the beginning of the radiation and matter-dominated epochs. In
accordance with Eq.~(\ref{t-H-relation1}), the obvious choices are
$b\Z{1}=b\Z{2}=0$, since during both matter and radiation
dominated epochs $t H \approx {\rm const}$. The requirements
$\Omega\Z{q} (1~{\rm MeV})< 0.1$ during big bang nucleosynthesis
(BBN) and $\Omega\Z{q}< 1$ during the matter-dominated universe
therefore imply that $n^2 < 1/40$ and $n<2/3$, respectively. This
result led us to conclude in~\cite{Ish:07e} that the agegraphic
dark energy with some fixed $n$ in (\ref{quantum-de}) is not a
viable alternative to concordance cosmology.

It would be possible to modify this outcome only by dropping one
or more premises of the standard model cosmology, such as, a
matter-dominated flat universe did not exist, which then tells
that the Einstein-de Sitter model is never realized truly. As an
illustrative example, one may consider the following modification
\begin{equation}\label{quantum-de-2}
\rho\Z{q} \propto \frac{1}{t^2\Z{P} (t+t\Z{1})^2} \equiv \frac{3
n^2\,m_{p}^2}{(t+ t\Z{1})^2},
\end{equation}
where $t\Z{1}$ is a constant with the dimension of time. In fact,
a solution of the above structure arises in almost all
scalar-tensor theories, e.g., with $V(q)\propto \e^{-\lambda\,
q/m\Z{P}}$ and $q(t)=(\lambda/2)\ln (t+t\Z{1})$ (see
e.g.~\cite{Ish:2005b}). In a standard approach, one normally sets
$t\Z{1}=0$ using the coordinate parameterization freedom of $t$,
with the assumption that such a shift in time only changes the
position of the big bang singularity. However, let us assume here
rather implicitly that no freedom was left so as to allow us to
set $t\Z{1}=0$; therefore, $t\Z{1}
> 0$ henceforth.

Then, typically, we may assume that $ t\Z{1}
> t\Z{0}$, where $t\Z{0}$ is the present age of the
universe. From the definition $\Omega\Z{q}\equiv n^2 /
[(t+t\Z{1})^2 H^2]$, we obtain
\begin{equation}\label{nonzero-delta}
\frac{n}{\sqrt{\Omega\Z{q}}}= (t+ t\Z{1}) H = t H
\left(1+\frac{t\Z{1}}{t}\right).
\end{equation}
A comparison between Eqs.~(\ref{main-soln}) and
(\ref{nonzero-delta}) shows that $b\Z{1}$ and $b\Z{2}$ are
nonzero; more precisely,
$$
t\Z{1}= \frac{b\Z{1}}{a\Z{\rm r, ~ini}}=
\frac{b\Z{2}}{a^{3/2}\Z{\rm m, ~ini}}.
$$
The bound $\Omega\Z{q} (1~{\rm MeV})< 0.1$ during the big-bang
nucleosynthesis (BBN) epoch may be satisfied for $40 n^2 < (1+
t\Z{1}/t)^2$. Next, consider that, at present, $t\Z{1}\equiv
2.33\, t\Z{0}$, $t\Z{0} \sim H\Z{0}^{-1}$ and $\Omega\Z{q}=0.73$.
This yields $w\Z{q}= -0.8$. That means, when the universe was a
half of the present age, $t/2 \sim \frac{3}{14}\, t\Z{1} \sim 6.8
~{\rm Gyrs}$ (approximately when $z\sim 1$), one had $w\Z{q}\simeq
-0.82$ (assuming a matter-dominated universe with $t H=2/3$), but
$\Omega\Z{q}\simeq 0.57$. If such a variation in the dark energy
density fraction is allowed by observations, then the agegraphic
dark energy model, with the modification (\ref{nonzero-delta}),
may be consistent with the concordance cosmology.

On the other end, if $t\Z{1}\ll t$ holds, then during the
matter-dominated epoch to which the WMAP and supernovae
measurements are sensitive, one finds $\Omega\Z{q} \simeq 9 n^2/4$
with $a(t) \propto t^{2/3}$, in which case, one obviously requires
$|n|<2/3$ during the matter-dominated epoch.

\subsection{Dark energy measured by a conformal time}

Next, let us consider another model of dynamical dark energy
proposed by Cai and Wei~\cite{Wei:07a}. In this proposal, one
takes the present-day dark energy density to be
\begin{equation}\label{quantum-CC-eta}
\rho\Z{q} \equiv \rho\Z{\Lambda} \propto \frac{1}{l\Z{P}^2\eta^2}
\equiv \frac{3 \alpha^2\,m\Z{P}^2}{\eta^2},
\end{equation}
where the numerical factor $3 \alpha^2$ is introduced for
convenience and $\eta$ is the conformal time
\begin{equation}\label{eta-time}
\eta= \int \frac{dt}{a}=  \int (aH)^{-1}\,d\ln a
\end{equation}
Differentiating Eq.~(\ref{eta-time}) with respect to $\ln {a}$,
one finds
\begin{equation}
\frac{d \eta}{d\ln a}= \frac{1}{a H}.
\end{equation}
Further, from the definition
\begin{equation}\label{defOmegaq}
\Omega\Z{q}\equiv \frac{\rho\Z{q}}{3 m\Z{P}^2
H^2}=\frac{\alpha^2}{\eta^2 H^2},
\end{equation}
we find
\begin{equation}\label{t-H-relation}
\eta H = \frac{\alpha}{\sqrt{\Omega\Z{q}}}
\end{equation}
Differentiating Eq.~(\ref{defOmegaq}) with respect to $\ln a$, we
obtain
\begin{equation}\label{constr-conformal}
\Omega\Z{q}^\prime +2 \varepsilon\Omega\Z{q} + \frac{2}{\eta
H}\,\e^{-\ln a} \Omega\Z{q} =0.
\end{equation}
Although $\alpha$ can take either sign, for a reason to be
explained, we shall normally take $\alpha< 0$; the choice for the
sign of $\alpha$ is actually linked to the choice of sign in
$d\eta \equiv \pm a\, dt$.

By Eq.~(\ref{quantum-CC-eta}) one can perhaps understand that the
universe starts out with zero vacuum energy, near the big bang,
since $\eta\to - \infty$. This may not look very physical from the
viewpoint that in almost all scalar field cosmologies the energy
of the vacuum or potential energy might drop sharply during
various phase transitions in the early universe. Nevertheless, the
magnitude of the present-day dark energy density determined by
Eq.~(\ref{quantum-CC-eta}) may be consistent with the cosmological
observations, for $|\alpha|
> 2.6$. In such a context, one should perhaps seek a dark energy
that behaves very differently than the standard scalar field
potential.

In fact, Eq.~(\ref{quantum-CC-eta}) draws some parallels with the
known example of quintessential potential, $V(q)\propto q^{-2}$.
It is generally expected that
\begin{equation}
\frac{\rho\Z{q}}{3} = \frac{1}{6}\, \dot{q}^2 + \frac{V(q)}{3}
\equiv \frac{\alpha^2 m\Z{P}^2}{\eta^2}.
\end{equation} In the limit $\dot{q}^2 \ll V(q)$, or simply that
$V(q)\propto \dot{q}^2$, we get
\begin{equation}
q^2 \propto \eta^2
\end{equation}
The limit of conformal time is $\eta\in (-\infty, 0)$; this then
translates to the condition that $|q|\to \infty $ near the big
bang, where $|\eta|\to \infty$, while $q\to 0$ in the asymptotic
future, $\eta\to 0$.

\section{Non-interacting dark energy, $\widetilde{Q}=0$}

Let us first consider the case $\widetilde{Q}=0$. From
Eq.~(\ref{EoM2}), we then get
\begin{equation}\label{wq-EoS1}
 w\Z{q} =
-1-\frac{1}{3}\frac{\Omega\Z{q}^\prime}{\Omega\Z{q}}-\frac{2\varepsilon}{3}.
\end{equation}
Comparing Eqs.~(\ref{constr-conformal}) and (\ref{EoM2}) we get
\begin{equation}\label{wq-constr}
w_{q}=-1+ \frac{2}{3\alpha} \,\e^{-\ln{a}} \,\sqrt{\Omega_{q}}.
\end{equation}
Equating Eqs.~(\ref{wq-EoS1}) and (\ref{wq-constr}), and then
solving for $\Omega\Z{q}$, we find
\begin{equation}\label{sol-in-epsilon}
\frac{1}{\sqrt{\Omega\Z{q}}}=\frac{c\, \alpha + \int \e^{-\ln a}
\, \left(\e^{-\int \varepsilon\, d\ln{a}}\right)
d\ln{a}}{\alpha\,\e^{-\int\varepsilon \ln a}}=H \left(c
+\frac{1}{\alpha} \int (a^2 H)^{-1}\,d{a}\right),
\end{equation}
where $c$ is an integration constant. In the discussion below we
often use the relation $\e^{\ln a}=(1+z)^{-1}$, where $z$ is the
redshift parameter, so that $a (z=0)\equiv a\Z{0}=1$.

Equation~(\ref{sol-in-epsilon}) gives rise to
\begin{equation}\label{main-soln-1}
\frac{1}{\sqrt{\Omega\Z{q}}}=  \left\{\begin{array}{l}
(\alpha\,a)^{-1}+b\Z{1}\,a^{-2} \qquad ({\rm RD}, \quad a\propto t^{1/2}), \\ \\
 2 (\alpha\,a)^{-1}+b\Z{2}\,a^{-3/2} \qquad ({\rm MD},
  \quad a\propto t^{2/3}),
\end{array} \right.
\end{equation}
where $b\Z{1}$, $b\Z{2}$ are integration constants. With the
choice $b\Z{1}=0=b\Z{2}$~\footnote{Or simply that $b\Z{1}\ll
a\Z{r, e}$ and $b\Z{2} \ll a\Z{m, e}$, where $a\Z{r, e}$ and
$a\Z{m, e}$ are the scale factors at the end of radiation and
matter-dominated epochs.} one finds $w\Z{q}=-1/3$ (RD) or
$w\Z{q}=-2/3$ (MD). Moreover, $\rho\Z{q} \propto 1/a^2$ (RD) or
$\rho\Z{q}\propto 1/a$ (MD). However, especially, with
$b\Z{i}>0$~\footnote{Unlike for the model in~\cite{Cai:07a},
$\Omega\Z{q}$ can be varying even deep into the matter-dominated
universe ($tH=2/3$) since $\eta H\ne {\rm const}$.}, one finds $-1
< w\Z{q} < -1/3$ (RD) or $-1 < w\Z{q} < -2/3$ (MD). If the
integration constants $b\Z{1}$, $b\Z{2}$ can be large, namely
$b\Z{1}\gg a\Z{r, e}$ and $b\Z{2}\gg a\Z{m, e}^{1/2}$, then during
both the RD and MD epochs, $\rho\Z{q}\propto {\rm const}$, which
mimics the case of a cosmological constant term.

Next, we consider a power-law expansion $a(t)\equiv [c\Z{0}
t+t\Z{1} ]^m$, with an arbitrary $m$. We then find
\begin{equation}\label{main-sol1}
\frac{1}{\sqrt{\Omega\Z{q}}}=  \left\{\begin{array}{l}
- \, \frac{m}{m-1}\,  \left( \alpha \,a\right)^{-1} + c\Z{1}\, a^{-1/m} \quad (m \ne  1), \\ \\
\ln a \left(\alpha\,a\right)^{-1} + c\Z{2}\,a^{-1} \qquad (m =1),
\end{array} \right.
\end{equation}
where $c\Z{1}$, $c\Z{2}$ are integration constants. Notice that,
for the branch $m>1 $, a physical solution may require $ \alpha$
to be negative, otherwise the quantity $\sqrt{\Omega\Z{q}}$
diverges at some stage of cosmic evolution, for $c\Z{1} >0$. Of
course, the choice $\alpha>0$ and $c\Z{1} <0$ is also allowed. In
either case, $w\Z{q} < -1$, since $\frac{1}{\alpha}
\sqrt{\Omega\Z{q}} < 0$.

\begin{figure}[ht]
\begin{center}
\hskip-0.4cm
\epsfig{figure=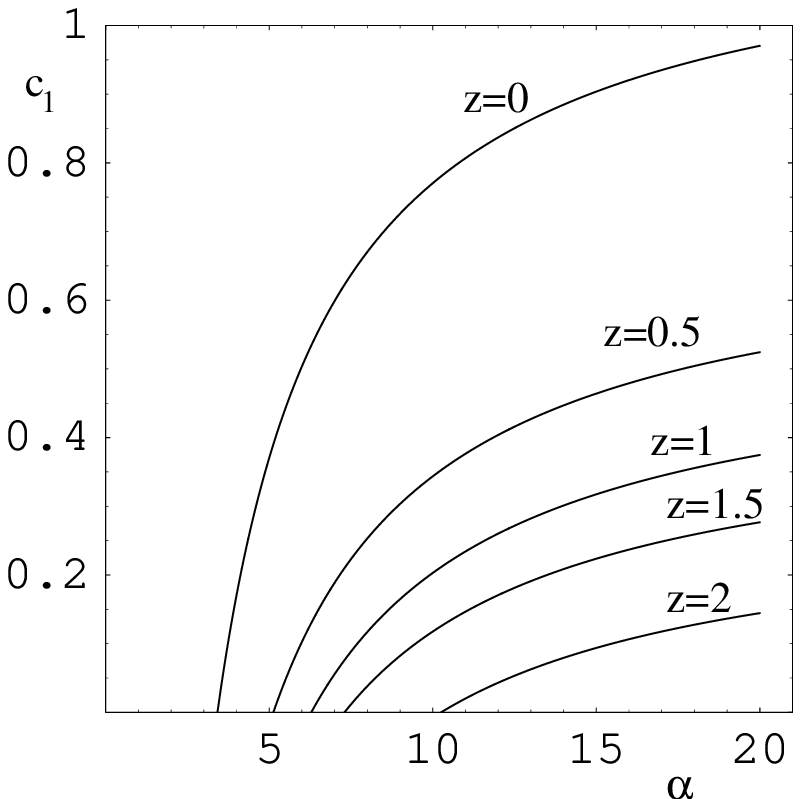,height=2.2in,width=3.1in}
\hskip0.5cm
\epsfig{figure=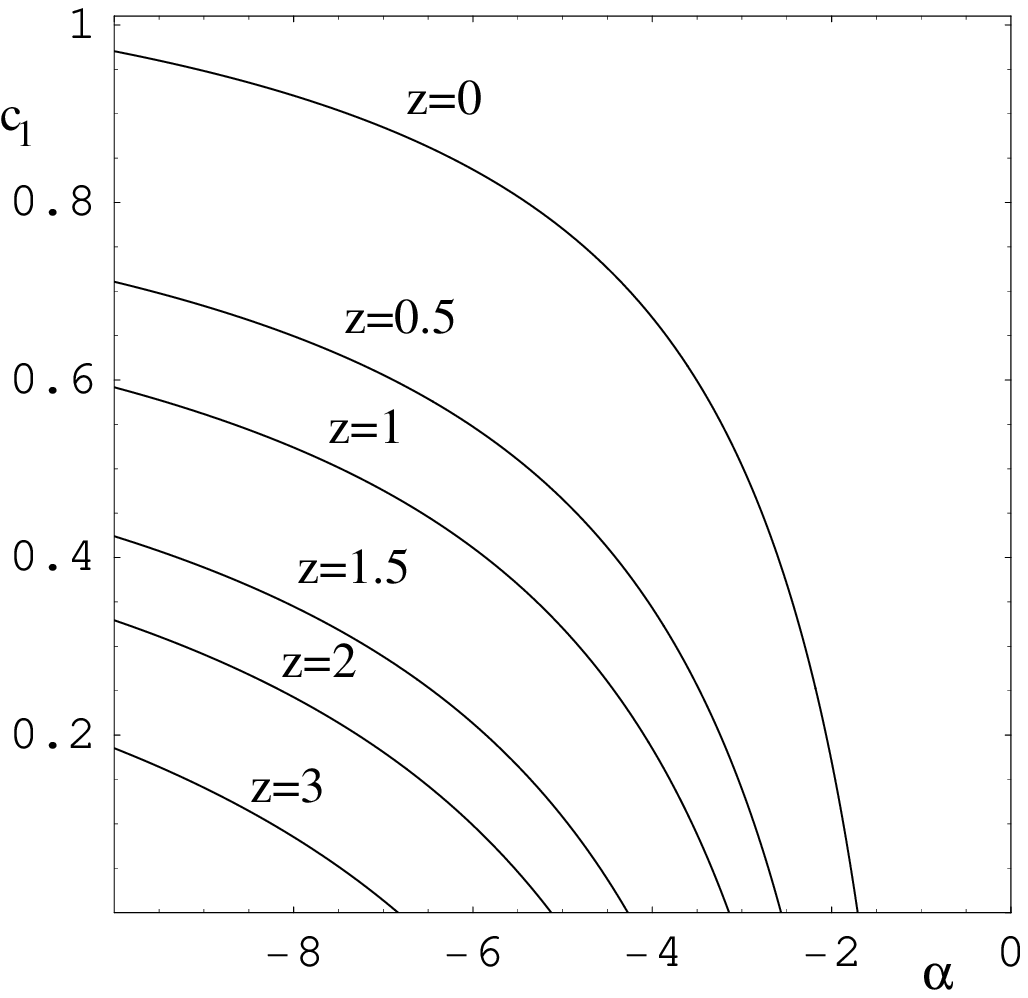,height=2.2in,width=3.1in}
\end{center}
\caption{The contour lines that give rise to dark energy density
parameter $\Omega\Z{q}=0.73$, with $m=0.8$ (left plot) and $m=2$
(right plot). $z$ is the redshift parameter defined via
$z=\e^{-\ln a}-1$. }\label{Fig1}
\end{figure}

A somewhat amusing result is, however, that one can adjust the
parameters $c\Z{1}$ and $\alpha$ such that $\Omega\Z{q}\simeq
0.73$ and $w\Z{q}< -1$ even for $m < 1$ (or $\varepsilon<-1$), in
which case the universe would be decelerating (cf
Fig.~\ref{Fig1}).

For solving the system of equations (\ref{EoM1})-(\ref{EoM3}),
analytically, one should perhaps make one or more simplifying
assumptions. It is worth noting that most of the radiation energy
in the present universe is in the cosmic microwave background,
which makes up a fraction of roughly $5\times 10^{-5}$ of the
total density of the universe. For this reason, let us make the
assumption that the matter is described by a pressureless
(non-relativistic) perfect fluid, i.e. $w\Z{m}\simeq 0$. We then
get
\begin{eqnarray}
\varepsilon &=& -\,
\frac{\Omega\Z{m}^\prime}{2\Omega\Z{m}}-\frac{3}{2} =-\,
\frac{\Omega\Z{r}^\prime}{2\Omega\Z{r}}-2,\nonumber \\
\Omega\Z{q} &=& 1- \left(1+ C\, \e^{\ln a}\right)
\Omega\Z{r},\qquad  \Omega\Z{m} = \Omega\Z{r} \,C\, \e^{\ln a}.
\end{eqnarray}
The numerical constant $C$ may be fixed using observational
inputs. Ideally, $\Omega\Z{q}\simeq 0.73$ and $\Omega\Z{m}\simeq
0.27$ at the present epoch ($a\simeq 1$) imply that $C\simeq
5400$. The matter-radiation equality epoch, $\Omega\Z{r}\simeq
\Omega\Z{m}$, then corresponds to the scale factor
$a\simeq1.85\times 10^{-4} a\Z{0}$ (where $a\Z{0}$ is the present
value of $a$). That means, the universe may have experienced about
$8.6$ e-folds of expansion since the epoch of matter-radiation
equality. This result is almost a model independent outcome, as
long as $w\Z{m}\simeq 0$ holds during the matter dominance.

\begin{figure}[ht]
\begin{center}
\hskip-0.4cm
\epsfig{figure=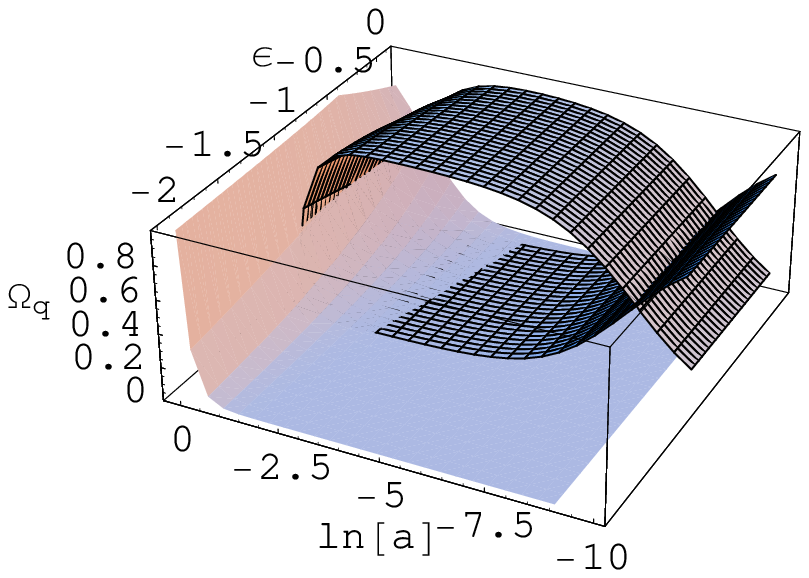,height=2.2in,width=3.3in}
\hskip0.2cm
\epsfig{figure=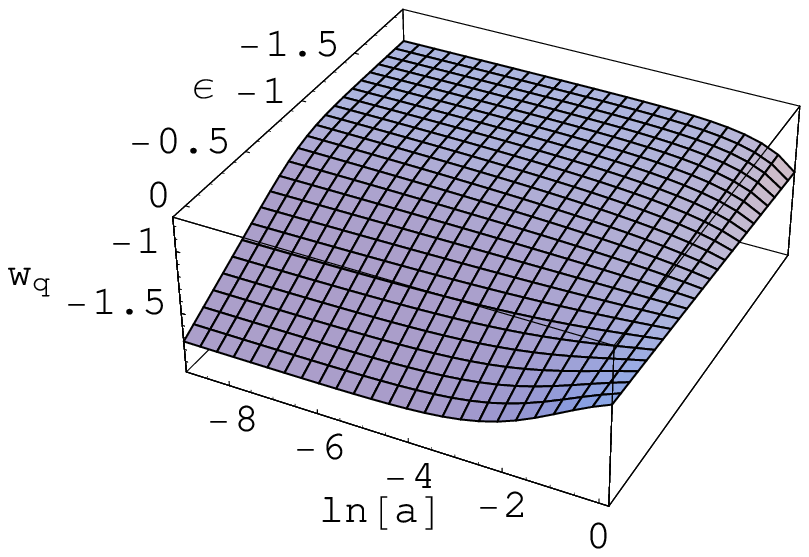,height=2.1in,width=3.1in}
\end{center}
\caption{The density parameters $\Omega\Z{m}$, $\Omega\Z{r}$,
$\Omega\Z{q}$ (from top to bottom) and the dark energy EoS
parameter $w\Z{q}$. We have taken $\alpha=- 2.7$. }\label{Fig2}
\end{figure}

\begin{figure}[ht]
\begin{center}
\hskip-0.5cm
\epsfig{figure=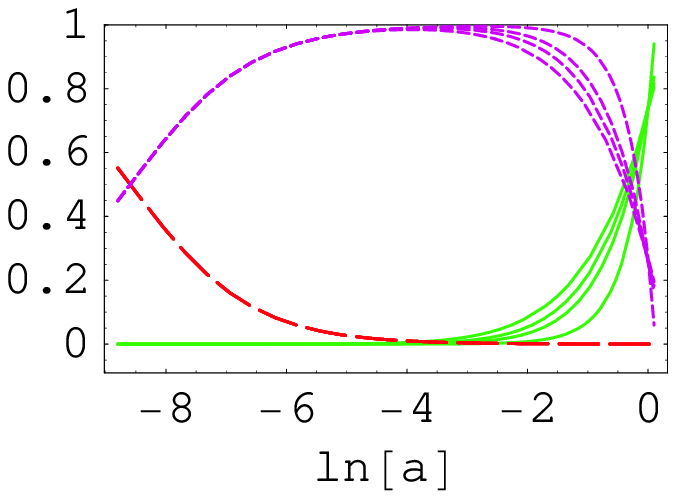,height=2.2in,width=3.6in}
\hskip0.1cm
\epsfig{figure=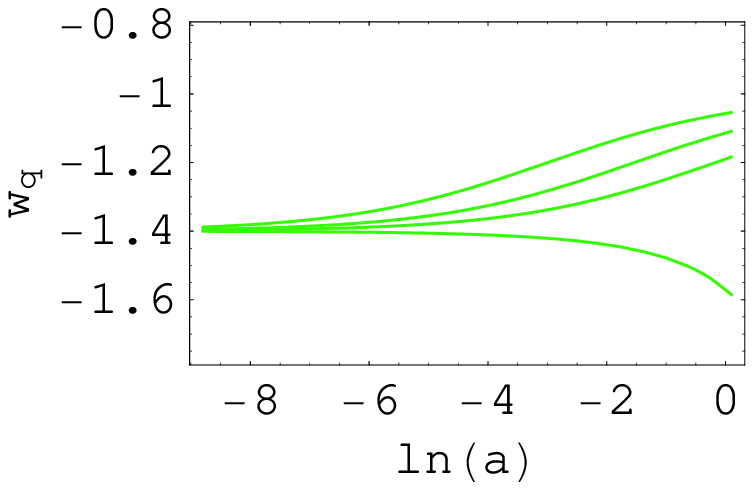,height=2.2in,width=3.1in}
\end{center}
\caption{The density parameters $\Omega\Z{m}$ (short-dashed,
pink), $\Omega\Z{r}$ (long-dashed, red) and $\Omega\Z{q}$ (solid,
green) with $\varepsilon=-0.4$ and $|\alpha|=1, 3, 5, 10$ (top to
bottom for $\Omega\Z{m}$ and $w\Z{q}$, while opposite for
$\Omega\Z{q}$).}\label{Fig3}
\end{figure}

Let us choose the integration constant $c\Z{1}$ in
(\ref{main-sol1}) such that $\Omega\Z{q}\simeq 0.73$ at present,
$a=a\Z{0}= 1$. This yields
\begin{equation} c\Z{1} =1.17+ \frac{1}{\alpha(1+\varepsilon)}\quad
(\alpha<0),\quad {\rm or} \qquad c\Z{1} =- 1.17+
\frac{1}{\alpha(1+\varepsilon)}\quad (\alpha> 0).
\end{equation}
By satisfying either of these conditions one gets
$\Omega\Z{q}=0.73$ at $a=1$, for any value of $\alpha$.
Figures~\ref{Fig2} and \ref{Fig3} show the evolution of density
parameters $\Omega\Z{r}$, $\Omega\Z{m}$, $\Omega\Z{q}$, and the
equation of state $w\Z{q}$.

\begin{figure}[ht]
\begin{center}
\hskip-0.5cm
\epsfig{figure=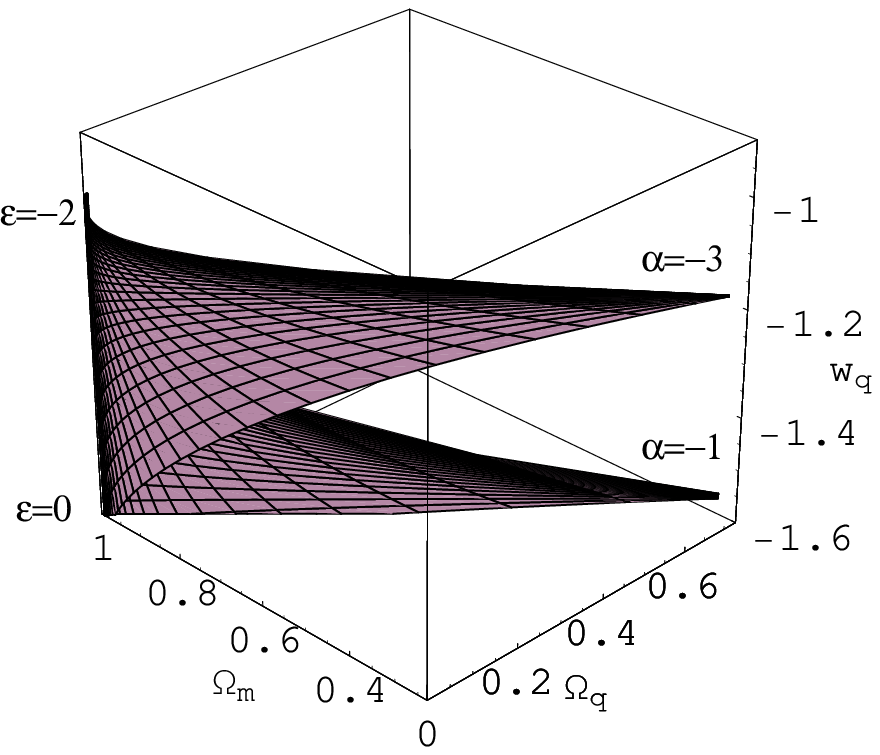,height=2.7in,width=3in}\hskip0.4cm
\epsfig{figure=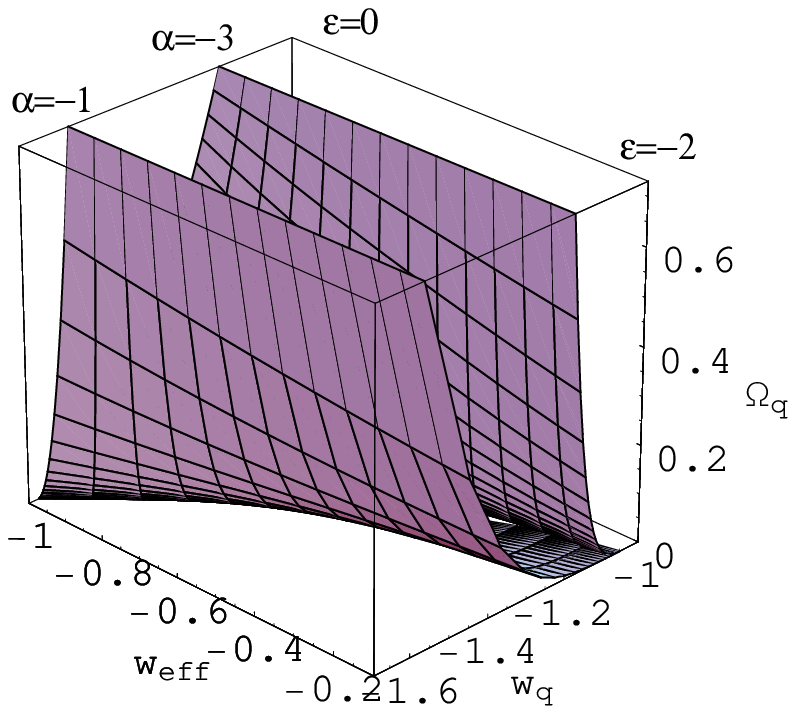,height=2.7in,width=3in}
\end{center}
\caption{A parametric 3D plot in the range $-\, 4\le \ln a \le 0$,
where $\Omega\Z{m}+\Omega\Z{q} \simeq 1$, with $|\alpha|=1, 3$. As
seen in the plots, the dark energy equation of state $w\Z{q}$
depends in the past on the acceleration parameter $\varepsilon$,
but for $\Omega\Z{q} \gtrsim \Omega\Z{m}+\Omega\Z{r}$, it is
highly dependent on the choice of the parameter
$\alpha$.}\label{Fig4}
\end{figure}

The model of dark energy in~\cite{Wei:07a} possesses some distinct
features as compared to a simpler model in~\cite{Cai:07a}.
Notably, due to the presence of the factor $\e^{-\ln a}$ in
Eq.~(\ref{wq-constr}), the dark energy equation of state parameter
$w\Z{q}$ does not behave, even in the limit $\Omega\Z{q}\to 0$, as
that for a cosmological constant term, for which $w\Z\Lambda=-1$;
the EoS $w\Z{q}$ rather depends on the acceleration parameter
$\varepsilon$, as is clearly seen from Fig.~\ref{Fig4}. Anyhow, in
the case $\widetilde{Q}=0$, there is no solution for which
$\Omega\Z{q}\simeq 0.73$ and $w\Z{q}\sim -1$ unless that
$|\alpha|\to \infty$.

In order get a cosmological evolution with $w\Z{q}\sim -1$, as
required for the best-fit concordance model cosmology, one should
perhaps consider the case $\widetilde{Q}\ne 0$. A mechanism that
works only for $\widetilde{Q}=0$ solves nobody's problem; it
perhaps only represents our ignorance about a universal coupling
between a fundamental scalar (or dark energy) field and the
ordinary (baryonic and dark) matter.

\section{Interacting dark energy, $\widetilde{Q}\ne 0$}

The cosmological observations have provided a strong evidence that
the current expansion of the universe is accelerating. In the
following discussion, we therefore assume that $\varepsilon>-1$.
Especially, in the case $\varepsilon\simeq {\rm const}$, because
of the constraint (\ref{quantum-CC-eta}), the following particular
solution
\begin{equation}\label{sol-non-zero-Q}
\frac{1}{\alpha} \sqrt{\Omega\Z{q}}= (1+\varepsilon) \left[ \alpha
c\Z{1} (1+\varepsilon) \,e^{\varepsilon \ln a}- \e^{-\,\ln
a}\right]^{-1},
\end{equation}
where $c\Z{1} $ is an integration constant, is also a viable
solution to the system of equations (\ref{EoM1})-(\ref{EoM3}),
with $\widetilde{Q}\ne 0$. However, as a notable difference, the
dark energy equation of state is now given by
\begin{equation}\label{new-wq-constr}
w\Z{q}=-1+ \frac{2}{3\alpha}\,\e^{-\ln a}
\sqrt{\Omega\Z{q}}-\frac{\widetilde{Q}}{3\Omega\Z{q}}.
\end{equation}
We shall normally take $\alpha c\Z{1} <0$, otherwise the quantity
$\Omega\Z{q}$ diverges at some stage of evolution. As a
consequence, the quantity $\alpha^{-1} \sqrt{\Omega\Z{q}}$ remains
negative. Eq.~(\ref{new-wq-constr}) then shows that it is possible
to get $w\Z{q}\simeq -1$, given that $\widetilde{Q}<0$. This is a
viable scenario.

\begin{figure}[ht]
\begin{center}
\hskip-0.5cm
\epsfig{figure=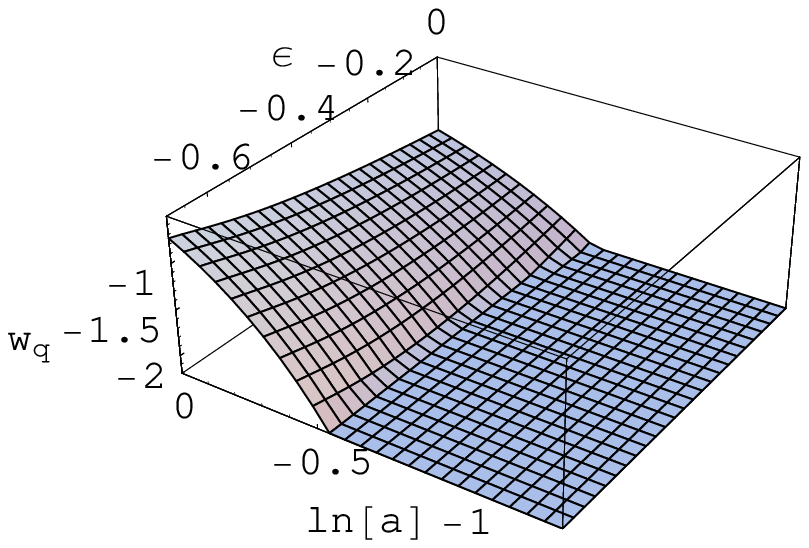,height=2.7in,width=3in}\hskip0.4cm
\epsfig{figure=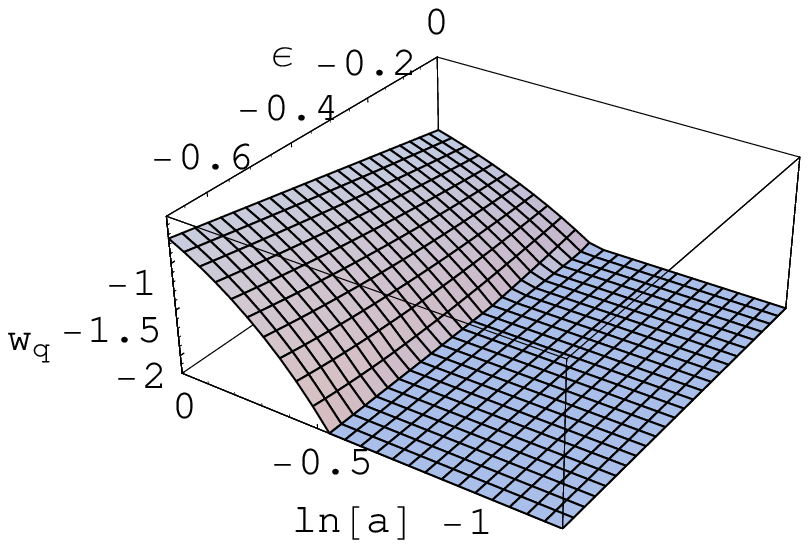,height=2.7in,width=3in}
\end{center}
\caption{The EoS parameter $w\Z{q}$ as in
Eq.~(\ref{new-wq-constr}) (left plot) and in
Eq.~(\ref{wq-reduced}) (right plot). We have taken here
$|\alpha|=2.7$ as suggested in~\cite{Wei:2007xu}.}\label{Fig5}
\end{figure}

To proceed analytically, let us assume that $w\Z{m}= 0$. Then, the
coupling $\widetilde{Q}$ is given by~\footnote{The form of
matter-scalar coupling that we consider in this paper precisely
follows a canonical kinetic structure of the theory determined by
the actions (\ref{action})-(\ref{matter-scalar}). A different
functional form for the scalar-matter coupling used, for example,
in~\cite{Bin-Wang}, namely $\widetilde{Q} \propto \Omega\Z{q}$,
may lead to a somewhat different result than found here.}
\begin{equation}
\widetilde{Q} \equiv  q^{\,\prime} \alpha\Z{q} \Omega\Z{m} =
\Omega\Z{m}^\prime+ 2\varepsilon \Omega\Z{m} +3 \Omega\Z{m}.
\end{equation}
One also notes that, with $w\Z{m}= 0$, the Friedmann equation
$\Omega\Z{\rm tot}=1$ gives rise to
\begin{eqnarray}
\Omega\Z{q} = 1- \left(1+ C\, \e^{\ln a}\right) \Omega\Z{r},\qquad
\Omega\Z{m} = \Omega\Z{r} \,C\, \e^{\ln a},
\end{eqnarray}
where $C$ is an integration constant. From Eq.~(\ref{de-EoS}), we
then get
\begin{equation}\label{wq-reduced}
w\Z{q} = \frac{2\varepsilon+3}{3\Omega\Z{q}},
\end{equation}
which is a valid approximation as long as $\Omega\Z{r}\ll 1$ and
$w\Z{m}\simeq 0$. As one would expect, the results coming from the
above two expressions for $w\Z{q}$, i.e.
Eqs.~(\ref{new-wq-constr}) and (\ref{wq-reduced}), agree at low
redshifts, that is, for $\ln a\lesssim 0$, see Fig~\ref{Fig5}.
This agreement is better for $|\alpha| \gg 1$, in which case
$\Omega\Z{q}$ overtakes $\Omega\Z{m}$ only at a slow rate.

One of the undesirable features of the model in~\cite{Wei:07a} is
that, as we go to higher redshifts, $\ln a\ll 0$, the coupling
$|\widetilde{Q}|$ decreases at a slower rate than the dark energy
density fraction $\Omega\Z{q}$, thereby leading to a divergent
$w\Z{q}\equiv p\Z{q}/\rho\Z{q}$ and/or a negative value for the
squared speed of sound, $v^2\equiv
dp\Z{q}/d\rho\Z{q}$~\footnote{This result is perhaps consistent
with the findings in a recent paper~\cite{Kim:2007iv}.}. A
possible resolution of this problem is to allow a much larger
value for $\alpha$ in the past, i.e. $|\alpha|\gg
1$~\footnote{This is opposite of that in the agegraphic dark
energy model~\cite{Cai:07a}, since $\eta\to -\infty$ in the
infinite past, whereas $t\to 0$ in the early universe.}. That
means, a dark energy model with some fixed value of $\alpha$ can
hardly explain most cosmological properties of our universe that
we observe. This is analogous to a situation in a standard scalar
field cosmology with a simple exponential potential $V\propto
\e^{-\lambda (\phi/m\Z{P})}$, having a constant slope parameter,
$\lambda={\rm const}$ (see, e.g.~\cite{Ish:2006ip} for a related
discussion).

While the assumption of power-law expansion of the scale factor
can be relaxed, e.g., during a transition from matter-dominance to
dark energy dominance, we do not expect it to greatly alter our
results.

\section{The holographic dark energy at a glance}

Some of the above difficulties may not arise in the model of
``holographic" dark energy proposed by Li and others. In this
model, the vacuum energy density is given by
\begin{equation}
\rho\Z{q}\equiv \rho\Z{\Lambda}=\frac{3 c^2 m\Z{P}^2}{R\Z{h}^2}.
\end{equation}
where \begin{equation}\label{event-horizon} R\Z{h}  \equiv a
\int_t^\infty \frac{dt_*}{a(t_*)} = a \int_{x}^\infty \frac{d
x}{H\, a}= \pm \frac{1}{H}\,\frac{c}{\sqrt{\Omega\Z{q}}}
\end{equation}
is the proper size of the future event horizon and $x\equiv \ln
a$. The last term in (\ref{event-horizon}) follows from the
definition $\Omega\Z{q}\equiv c^2/ H^2 R\Z{h}^2$. The analogue of
the constraint equation (\ref{constr-conformal}) is
\begin{equation}\label{constr1-Li}
\Omega\Z{q}^\prime +2 \varepsilon\Omega\Z{q} + 2\Omega\Z{q}
\left(1-\frac{1}{R\Z{h} H}\right) =0.
\end{equation}
For $c>0$, one takes the positive sign in
Eq.~(\ref{event-horizon}), so that $R\Z{h} H>0$.

In the absence of interaction between the $q$-field and matter, so
$\widetilde{Q}=0$, from Eqs.~(\ref{constr1-Li}) and (\ref{EoM2}),
we find
\begin{equation}\label{wq-constr-Li}
w_{q}=-\frac{1}{3} -\frac{2}{3 } \,\frac{1}{R\Z{h} H}=
-\frac{1}{3}-\frac{2}{3 c}\, \sqrt{\Omega\Z{q}}.
\end{equation}
In particular, for the power-law expansion $a=[c\Z{0} t+t\Z{1}
]^m$, the explicit solution is given by
\begin{equation}\label{main-sol1-Li}
\frac{1}{\sqrt{\Omega\Z{q}}}=  \left\{\begin{array}{l}
\frac{m}{(m-1) c} + c\Z{1} \,a^{(m-1)/m} \quad (m \ne  1), \\ \\
-\frac{1}{c}\,\ln a+ c\Z{2} \qquad (m =1),
\end{array} \right.
\end{equation}
where $c\Z{1}$ and $c\Z{2}$ are integration constants. Therefore,
by choosing
\begin{equation}\label{main-sol2-Li}
c\Z{1} \gg  \left\{\begin{array}{l} \frac{1}{c}\, a\Z{r, e} \quad
(a\propto t^{1/2},\quad {\rm RD}),\\ \\
\frac{2}{c}\, a\Z{m, e}^{1/2} \quad (a\propto t^{2/3},\quad {\rm
MD}),
\end{array} \right.
\end{equation}
where $a\Z{r, e}$ and $a\Z{m, e}$ are the scale factors at the end
of radiation and matter-dominated epochs, one finds $\Omega\Z{q}
\propto a$ (MD) and $\Omega\Z{q} \propto a^2$ (RD). This then
implies that during both the MD and RD epochs, the holographic
dark energy density scales as $\rho\Z{q} \propto 1/a^2$. It is
thus conceivable that the dark energy density overtakes both the
radiation and matter energy densities at some stage of cosmic
evolution since $\rho\Z{r}\propto 1/a^4$ and $\rho\Z{m}\propto
1/a^3$. The nucleosynthesis bound $\Omega\Z{q} (1 {\rm MeV})< 0.1$
may also be satisfied for almost any value of $c$, although $c<
1.17$ may be required to get $w\Z{q} < -0.82$ with the input
$\Omega\Z{q}\simeq 0.73$ at present. The constraint $c\ge
\sqrt{\Omega\Z{q}}$ may also be imposed by demanding that the de
Sitter entropy, $S\equiv A/4G_N=\pi m\Z{P}^2 R\Z{h}^2$ does not
decrease, that is $\dot{R}\Z{h}= -1+ c/\sqrt{\Omega\Z{q}}> 0$. A
detailed analysis with $\widetilde{Q}\ne 0$ appears elsewhere.

\section{Discussions}

Dynamical dark energy models with the vacuum energy density
$\rho\Z{q} \propto 1/t^2$ may lead to some undesirable features,
especially, during the matter and radiation-dominated epochs,
since $\rho\Z{q}\propto 1/a^3$ (MD) and $\rho\Z{q}\propto 1/a^4$
(RD). This rules out, for instance, a transition from
matter-dominance to dark energy-dominance, unless that the
late-time acceleration arises due to some other dynamics, e.g., a
nontrivial growing interaction between the $q$-field and matter.
This situation is improved by assuming that $\rho\Z{q}\propto
1/(t+t\Z{1})^2$, with $t\Z{1} \gtrsim t\Z{\rm present}\equiv
t\Z{0}$, as we discussed above.

The model of conformal dark energy proposed by Cai and
Wei~\cite{Wei:07a} may be consistent with quantum kinematics, in
the sense that the uncertainty relation (or the second law of
thermodynamics, in an equivalent form) is obeyed. Also, the model
does not suffer from the problem of causality, unlike the
holographic dark energy model, with $c<1$. Nevertheless, the
conformal dark energy model in~\cite{Wei:07a} has some undesirable
features, such as, in the presence of a nontrivial coupling
between the $q$-field and ordinary matter, the dark energy
equation of state parameter $w\Z{q}$ may diverge as higher
redshifts, thereby leading to a negative value for the squared
speed of sound, $v^2\equiv dp\Z{q}/d\rho\Z{q}$. The main reason
for this odd behaviour is that the dark energy density fraction
$\Omega\Z{q}$ varies (actually decreases) too fast in the past,
unless that $|\alpha|$ takes a value significantly larger than
unity, which is, however, not compatible with the epoch of matter
dominance, where $\Omega\Z{q} < 0.2$.

The other obvious drawback of the conformal dark energy model is
that it only provides a kinematic approach to dark energy, by
outlining a possible time decay of dark energy component, but the
model does not explain much about the dynamics, that is, the
origin or nature of dark energy. Both the conformal and
holographic dark energy models are interesting in the sense that
they satisfy some holographic entropy bounds (or laws of
thermodynamics, in equivalent forms). But they still raise some
other important concerns: Why quantum corrections to the vacuum
energy contribute to the present-day dark energy density ($\sim
10^{-12} ~ {\rm eV}^4$) dominantly, whereas many known
contributions to $\rho\Z{\Lambda}$, including the classical
effects of quantum fields, do not? and why it is comparable to the
energy density of matter today?

The holographic dark energy model is perhaps a step forward among
the recent attempts in probing a time-variation of dark energy
within the framework of quantum gravity, even though the model has
some pitfalls, such as, a semi-classical instability due to a
negative value for the squared of sound speed.

\medskip
{\sl Acknowledgements}: This research is supported by the FRST
Research Grant No. E5229 (New Zealand) and also by Elizabeth Ellen
Dalton Research Award (E5393).


\vskip -0.8cm \baselineskip 22pt
\begin{thebibliography}{99}
\itemsep 0pt

\bibitem{WMAP}
D.~N.~Spergel {\it et al.} [WMAP Collaboration],
  {\it First Year Wilkinson Microwave Anisotropy Probe (WMAP) Observations:
  Determination of Cosmological Parameters},
  Astrophys.\ J.\ Suppl.\ Ser.\ {\bf 148}, 175 (2003).

\bibitem{Linde:87} A. Linde, {\it Inflation and quantum
cosmology}, in 300 Years of Gravitation, edited by S.~W.~Hawking
and W.~Israel (Cambridge University Press, Cambridge, England,
1987), p. 604.

\bibitem{supernovae}
S. Perlmutter {\it et al.} (Supernova Cosmology Project
Collaboration), Astrophys. J. {\bf 517}, 565 (1999);
A.~G.~Reiss {\it et al.} (Supernova Search Team Collaboration),
Astrophys. J. {\bf 607}, 665 (2004).

\bibitem{Astier:05}
P.~Astier {\it et al.}  (The SNLS Collaboration),
  Astron.\ Astrophys.\  {\bf 447}, 31 (2006)
  [astro-ph/0510447].


\bibitem{Tegmark:SDSS}
M.~Tegmark {\it et al.} (SDSS Collaboration), Phys. Rev. D {\bf
69}, 103501 (2004). 

\bibitem{Spergel:06}
  D.~N.~Spergel {\it et al.}  (WMAP Collaboration),
  Astrophys.\ J.\ Suppl.\  {\bf 170}, 377 (2007)
  [astro-ph/0603449].


\bibitem{Weinberg:1988}
  S.~Weinberg,
  {\it The cosmological constant problem},
  Rev.\ Mod.\ Phys.\  {\bf 61}, 1 (1989).

\bibitem{Padmanabhan}
T.~Padmanabhan,
  {\it Cosmological constant: The weight of the vacuum},
  Phys.\ Rept.\  {\bf 380}, 235 (2003)
  [hep-th/0212290];
E.~J.~Copeland, M.~Sami and S.~Tsujikawa,
  {\it Dynamics of dark energy},
  Int.\ J.\ Mod.\ Phys.\  D {\bf 15}, 1753 (2006) [hep-th/0603057].

\bibitem{Ratra:88}
P.~J.~E.~Peebles and B.~Ratra, {\it Cosmology with a time variable
cosmological {`constant'}}, \ApJ{325}, L17 (1988);
  C.~Wetterich,
  {\it The Cosmon model for an asymptotically vanishing time dependent
  cosmological `constant'},
  Astron.\ Astrophys.\  {\bf 301}, 321 (1995)
  [hep-th/9408025].

\bibitem{Peebles:99a}
  P.~J.~E.~Peebles and A.~Vilenkin,
  {\it Quintessential inflation},
  Phys.\ Rev.\  D {\bf 59}, 063505 (1999).

\bibitem{Ish:07b}
  I.~P.~Neupane,
  {\it Model of Quintessential Inflation Consistent with
  Observations},
  arXiv:0706.2654 [hep-th].

\bibitem{Wei:07a}
  H.~Wei and R.~G.~Cai,
  {\sl A New Model of Agegraphic Dark Energy},
  arXiv:0708.0884.

\bibitem{Cohen:1998}
  A.~G.~Cohen, D.~B.~Kaplan and A.~E.~Nelson,
  {\it Effective field theory, black holes, and the cosmological
  constant},
  Phys.\ Rev.\ Lett.\  {\bf 82}, 4971 (1999)
  [hep-th/9803132];
S.~D.~Thomas,
  {\it Holography stabilizes the vacuum energy},
  Phys.\ Rev.\ Lett.\  {\bf 89} (2002) 081301.


\bibitem{Hsu:2004}
  S.~D.~H.~Hsu,
  {\it Entropy bounds and dark energy},
  Phys.\ Lett.\  B {\bf 594}, 13 (2004)
  [hep-th/0403052].

\bibitem{Li:2004}
  M.~Li,
  {\it A model of holographic dark energy},
  Phys.\ Lett.\  B {\bf 603}, 1 (2004)
  [hep-th/0403127].

\bibitem{Damour:1993id}
  T.~Damour and K.~Nordtvedt,
  {\it Tensor - scalar cosmological models and their relaxation toward general
  relativity},
  Phys.\ Rev.\  D {\bf 48}, 3436 (1993).

\bibitem{Ish:07e}
  I.~P.~Neupane,
  {\it A Note on Agegraphic Dark Energy},
  arXiv:0708.2910.

\bibitem{Ish:07a}
  B.~M.~Leith and I.~P.~Neupane,
  {\it Gauss-Bonnet cosmologies: crossing the phantom divide and the transition
  from matter dominance to dark energy},
  JCAP {\bf 0705}, 019 (2007) [hep-th/0702002].

\bibitem{Sahni}
V.~Sahni and A.~Starobinsky,
  {\it Reconstructing dark energy},
  Int.\ J.\ Mod.\ Phys.\  D {\bf 15}, 2105 (2006)
  [astro-ph/0610026].


\bibitem{Cai:07a}
  R.~G.~Cai,
  {\it A Dark Energy Model Characterized by the Age of the
  Universe},
  arXiv:0707.4049.

\bibitem{Sasakura}
N.~Sasakura, {\it An uncertainty relation of space-time}, Prog.\
Theor.\ Phys.\ {\bf 102}, 169 (1999) [hep-th/9903146].

\bibitem{Maziashvili}
  M.~Maziashvili,
  {\it Cosmological implications of Karolyhazy uncertainty
  relation},
  Phys.\ Lett.\  B {\bf 652}, 165 (2007)
  [arXiv:0705.0924].

\bibitem{Wu:07a}
  X.~Wu, Y.~Zhang, H.~Li, R.~G.~Cai and Z.~H.~Zhu,
  {\sl Observational Constraints on Agegraphic Dark Energy},
  arXiv:0708.0349.

\bibitem{Ish:2005b}
  I.~P.~Neupane and D.~L.~Wiltshire,
  {\it Cosmic acceleration from M theory on twisted spaces},
  Phys.\ Rev.\  D {\bf 72}, 083509 (2005)
  [arXiv:hep-th/0504135].

\bibitem{Bin-Wang}
  B.~Wang, Y.~g.~Gong and E.~Abdalla,
  {\it Transition of the dark energy equation of state in an interacting
  holographic dark energy model},
  Phys.\ Lett.\  B {\bf 624}, 141 (2005)
  [hep-th/0506069];
  H.~Wei and R.~G.~Cai,
  {\it Interacting Agegraphic Dark Energy},
  arXiv:0707.4052.

\bibitem{Kim:2007iv}
  K.~Y.~Kim, H.~W.~Lee and Y.~S.~Myung,
  {\it Instability of agegraphic dark energy models},
  arXiv:0709.2743.


\bibitem{Ish:2006ip}
  I.~P.~Neupane,
  {\it Towards inflation and accelerating cosmologies in string-generated gravity
  models},
  arXiv:hep-th/0605265.

\bibitem{Wei:2007xu}
  H.~Wei and R.~G.~Cai,
  {\it Cosmological Constraints on New Agegraphic Dark Energy},
  arXiv:0708.1894.



\end{thebibliography}
\end{document}